\documentclass[epj,final]{svjour}
\usepackage{amssymb}
\usepackage{epsfig}
\usepackage[latin1]{inputenc}
\usepackage[T1]{fontenc}

\renewcommand{\Re}{{\rm Re}} 
\renewcommand{\Im}{{\rm Im}} 
\begin{document}
\title{Euler-Heisenberg Lagrangian to all orders in the magnetic field
and the Chiral Magnetic Effect}
\medskip

\author{Simon Wolfgang Mages\inst{1}, Matthias Aicher\inst{1},
Andreas Sch\"afer\inst{1}}
%
%
\institute{\inst{1}
Institut f{\"u}r Theoretische Physik, Universit{\"a}t Regensburg,
D-93040 Regensburg, Germany}
%
\date{}
%
\abstract{
In high energy heavy ion collisions as well as in astrophysical objects 
like magnetars extreme magnetic field strengths are reached. 
Thus, there exists a need to calculate divers QED processes to all orders 
in the magnetic field. We calculate the vacuum polarization graph 
in second order of the electric field and all orders of the magnetic field
resulting in a generalization of the Euler-Heisenberg Lagrangian.
We perform the calculation in the effective Lagrangian approach of 
J. Schwinger as well as using modified Feynman rules. We find that 
both approaches give the same results provided that the different 
finite renormalization terms are taken into account.
Our results imply that any quantitative explanation of 
the recently proposed {\em Chiral Magnetic Effect} has to take 
'Strong QED' effects into account, because these corrections 
are huge. 
\PACS{
      {12.20.-m}{}   \and
      {12.20.Ds}{}   
     } 
} 

\authorrunning{Mages, Aicher, Sch\"afer}
\titlerunning{Euler-Heisenberg to all  orders and CME}

\maketitle
\section{Motivation and Introduction}
Long time ago J.~Schwinger derived in a seminal paper \cite{schwinger}
the effective non-linear Lagrangian for constant electric and magnetic 
fields in all orders. Later, the exact fermion propagator in a constant magnetic field
was derived based on this work \cite{FeynProp,ruder} and used to 
treat e.g. QED processes in magnetars 
\cite{Lai,Meszaros:1979xf,Herold1,Herold2,Heyl:2005an}. 
The latter is 
important, because the observed spectra are strongly affected by 
QED effects and the deduction of e.g. the magnetic field strength 
reached in these objects or the structure of the accretion column 
depends crucially on the quality of 
these calculations.\\ 
In heavy ion collisions even far stronger magnetic 
fields, of the order of  
$|\vec B| \sim (100 \mbox{ MeV})^2$ and above,
are generated for a short time period. Recently, the STAR 
experiment at BNL observed correlations between charged hadrons 
which can best be understood, if one assumes that topologically
non-trivial QCD effects allow for an effective, naively CP-odd coupling   
of the type $\vec E \cdot \vec B$, see \cite{cms1,cms2}. 
This hypothetical mechanism is called {\em Chiral Magnetic Effect} (CME).
In \cite{ba} it was argued that the electric field induced by any such 
effect should minimize the energy and thus is determined by the  
linear term from the coupling to $G^a_{\mu\nu}\tilde G^{a\mu\nu}$ 
and the field energy term, quadratic in $E$.
Therefore, whatever the precise nature of the CME might be, 
it will be affected strongly by pure QED effects which drastically 
change the electromagnetic field energy for such strong fields.  
Thus they will alter the magnitude of the induced electric field strength
and thus the size of the charged particle correlations.
These QED effects are even important for all charge correlations,  
independently of whether 
the CME is confirmed by future measurements or not.\\ 
In principle these calculations should take the detailed dynamics of 
heavy-ion collisions into account. In the present work, however, 
we only discuss the case of constant fields, which already involves 
some conceptual problems. \\
In the CME the induced electric fields are relatively weak, roughly 
of the order of $(10-20~{\rm MeV})^2$, such that it is sufficient to 
determine the contribution which is quadratic in $\vec E$ but 
includes all orders in $\vec B$.\\
The main technical problem we are facing is the following: 
Schwinger's elegant calculation is based on his 
proper time formalism, which from the very beginning expresses 
the effective Lagrangian in terms of the gauge invariant  
field strength tensor $F^{\mu\nu}$. For many dynamical applications one does 
need, however, the fermion propagator for which an explicit form
is given e.g. in \cite{ruder}. We, therefore, reproduced the Schwinger 
result also in that formalism, which was highly non-trivial and actually 
required a more careful definition of the exact fermion propagator than the one given in \cite{ruder}.\\
These technical problems are also reflected by the literature. In many
papers the effective action result of Schwinger is used to analyze 
specific situation beyond the weak field limit, e.g. \cite{Cho:2006cv}.
In others the problem is discussed that its expansion in powers of the fields 
leads to an asymptotic series, which is very difficult to handle
and discouraged some applications. However, it seems that this is 
not a fundamental problem but just an unlucky choice of expansion
as was shown e.g. in \cite{Cho:2000ei}. In this paper we will therefore 
avoid any expansion. This will lead us to expressions containing the 
$\psi$ function, which indeed has an asymptotic expansion involving 
Bernoulli numbers but has perfectly reasonable properties if not expanded.\\
In e.g. \cite{Beneventano} it was discussed with great clarity that finite 
regularisation terms have to be treated with care to avoid misinterpretations, 
but that they do not pose a problem of principle. In our calculation we are 
interested in higher order terms for which such problems do not occur 
and adopt for the leading terms just the standard results.\\  
The outline of this paper is as follows:
In section 2 we shall present the effective Lagrangian calculation within
the
Schwinger formalism and in section 3 we discuss the problems encountered 
when trying to do the same calculations with Feynman rules.
In section 4 we will conclude and discuss our findings in the context of  
the CME.

\section{The calculation using Schwinger's effective Lagrangian formalism }

\begin{figure}
	\centering
		\includegraphics[width=0.45\textwidth]{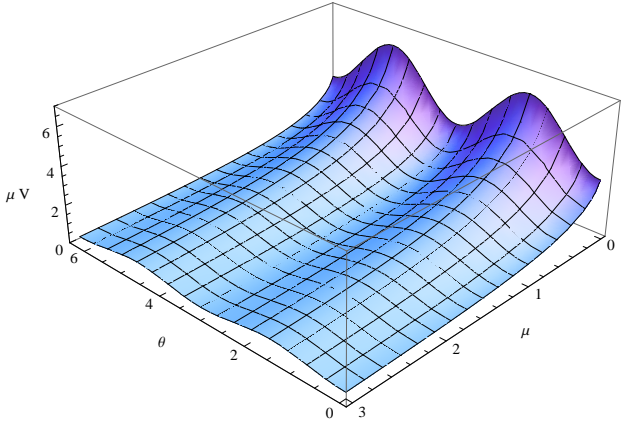}
	\caption{Potential energy correction in leading order of the electric field as a function of the angle between the electric and magnetic field $\Theta$ and $\mu=m^2/(eB)$; for better visualization of the behaviour at $\mu=0$ in this plot the potential is multiplied by $\mu$, and additionally normalised to the corresponding prefactor of the Euler Heisenberg lagrangian $e^2 E^2/(72 \pi^2)$.}
	\label{fig:3dThetasVar}
\end{figure}

\begin{figure}
	\centering
		\includegraphics[width=0.45\textwidth]{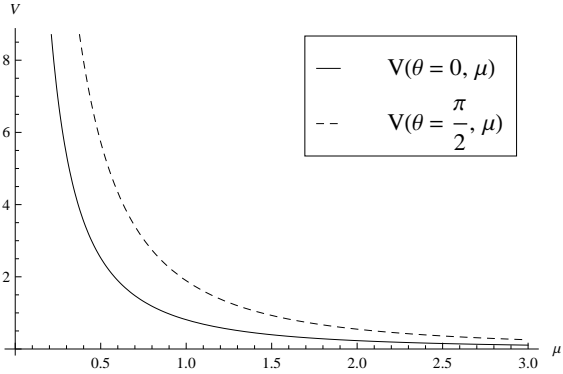}
	\caption{Potential energy correction in leading order of the electric field as a function of $\mu=m^2/(eB)$ for values of the angle between the electric and magnetic field $\Theta=0$ and $\Theta=\pi/2$; in this plot the potential is normalised to the corresponding prefactor of the Euler Heisenberg lagrangian $e^2 E^2/(72 \pi^2)$.}
	\label{fig:2dThetas}
\end{figure}

Schwinger's expression for the effective Lagrangian 
of constant electromagnetic fields reads \cite{schwinger}
\begin{eqnarray}
\label{schwinger}
{\cal L}^{(1)}&=&-\frac{1}{8\pi^2}\int_0^\infty ds s^{-3} \exp(-m^2s) 
\nonumber\\
&&\left[(es)^2 {\cal G} \frac{\Re \cosh(esX)}{\Im \cosh(esX)}-1\right],
\end{eqnarray}
where
\begin{eqnarray}
	X&=&(2({\cal F}+i{\cal G}))^{\frac{1}{2}}\\
	{\cal G}&=&\vec{E}\cdot\vec{B}\\
	{\cal F}&=&\frac{1}{2}(\vec{B}^2-\vec{E}^2)
\end{eqnarray}
and  $m$ is the mass of the considered Dirac field, here of the electron. 
With
\begin{eqnarray}
	B&:=&\left|\vec{B}\right|\\
	E&:=&\left|\vec{E}\right|\\
	EB\cos\Theta&:=&\vec{E}\cdot\vec{B}
\end{eqnarray}
the real and imaginary parts read
\begin{eqnarray}
\Re \cosh(esX)&=&\cosh(esB)\left(1-\frac{1}{2}(es)^2E^2\cos^2\Theta
\right.
\\
&-&\left. \frac{1}{2}\tanh(esB)es\frac{E^2}{B}\sin^2\Theta+O(E^4)\right)
\nonumber\\
\Im \cosh(esX)&=&\sinh(esB)esE\cos\Theta
\\
&\times&\left(1-\frac{1}{6}(es)^2(E\cos\Theta)^2\right.
\nonumber\\
&+&\frac{1}{2}E^2\sin^2\Theta
\left(\frac{1}{B^2}-\frac{es}{B}\coth(esB)\right)
\nonumber \\
&+& \left. O(E^4)\right)
\nonumber
\end{eqnarray}
and Eq.(\ref{schwinger}) simplifies to 
\begin{eqnarray}
{\cal L}^{(1)}&=&
-\frac{1}{8\pi^2}\int_0^\infty ds s^{-3} \exp(-m^2s) 
\nonumber \\
&\times& \left[ esB\coth(esB)\left(1-E^2\left( \frac{e^2s^2}{3} 
\cos^2\Theta  \right.\right.\right.\nonumber\\
&-&\sin^2\Theta\left(\frac{es}{2B}\left(-\tanh(esB)
+\coth(esB)\right)\right.
\nonumber \\
&-& \left.\left.\left.\left. \frac{1}{2B^2} \right) \right) +O(E^4)\right)-1\right].
\end{eqnarray}
For renormalisation one has to subtract the logarithmic divergence 
and one has to decide on the finite renormalisation one chooses. In principle one
could e.g. subtract the contribution for any fixed magnetic field 
$\vec B_0$. However, there is no good reason to introduce such an 
additional parameter. We follow 
Schwinger in subtracting the limiting case for vanishing magnetic field
\begin{eqnarray}
{\cal L}^{(1)}_{B=0}&=&-\frac{1}{8\pi^2}\int_0^\infty ds s^{-3} 
\exp(-m^2s)\frac{e^2s^2}{3}\left(B^2-E^2\right)
\nonumber \\
&+& O(E^4)
\end{eqnarray}
and thus obtain for the effective, renormalised Lagrangian
of second order in $\vec E$ and all orders in $\vec B$ 
\begin{eqnarray}
\label{renLag}
{\cal L}^{(1)}_{ren}&=& {\cal L}^{(1)}-{\cal L}^{(1)}_{B=0} 
\nonumber\\
&=&{\cal L}^B(B)+{\cal V}_{eff}(B,E)+{\cal V}^\Theta_{eff}(B,E,\sin\Theta)
\nonumber \\
&+&O(E^4)
\\
{\cal L}^B(B)&=&-\frac{1}{8\pi^2}\int_0^\infty ds s^{-3} \exp(-m^2s)
\nonumber \\
&\times& \left(esB\coth(esB)-1-\frac{e^2s^2}{3}B^2\right)\\
{\cal V}_{eff}(B,E)&=&
\frac{E^2e^2}{24\pi^2}\int_0^\infty ds s^{-1} 
\exp(-m^2s)
\nonumber \\
&\times&\Bigl(esB\coth(esB)-1\Bigr)\\
&=&
\frac{e^2E^2}{24\pi^2}\left(\log\left(\frac{\mu}{2}\right) - \Psi\left(\frac{\mu}{2}\right) 
- \frac {1}{\mu}\right)\\
\mu &=& \frac{m^2}{eB}\\
{\cal V}^\Theta_{eff}(B,E,\sin\Theta)&=&
\frac{E^2e^2}{16\pi^2}\sin^2\Theta\int_0^\infty ds s^{-1} \exp(-m^2s)\nonumber\\
&\times&\left(-\frac{1}{\sinh^2(esB)}+\frac{\coth(esB)}{esB}\right.
\nonumber \\
&-& \left.\frac{2}{3}esB\coth(esB)\right).
\end{eqnarray}
The sign convention was chosen such that the contribution of e.g. 
${\cal V}_{eff}$ to the energy density is
\begin{equation}
\label{eq:energy_density}
{\cal H}^{(1)}_{\rm ren} ~=~ E\frac{\partial {\cal L}^{(1)}_{\rm ren}}
{\partial E} -{\cal L}^{(1)}_{\rm ren}~=~{\cal V}_{eff} 
\end{equation}
The behaviour of the potential ${\cal V}_{eff}+{\cal V}^\Theta_{eff}$ is shown in Fig. \ref{fig:3dThetasVar} and Fig. \ref{fig:2dThetas}.
It is easy to verify that for small $\vec B$ fields one reproduces all contributions of second order in $E$ to the
Euler-Heisenberg Lagrangian
\begin{eqnarray}
\label{eq:euler_heisenberg}
{\cal L}^{(1)}_{ren}&=&
\frac{2\alpha^2}{45m^4}\left[\left(\vec{E}^2-\vec{B}^2\right)^2+
7\left(\vec{E}\cdot\vec{B}\right)^2\right]
+O\left(\left(\frac{eB}{m^2}\right)^6\right)
\nonumber \\
&+&O(E^4).
\end{eqnarray}
For the CME we are most interested in the case $\Theta=0$, i.e. in 
${\cal V}_{eff}(B,E)$. 
The extension of ${\cal L}^{(1)}_{ren}$ to higher orders in $E$ is quite straight forward. 
For the term of interest one gets to arbitrary order $N>1$ 
\begin{eqnarray}
\label{eq:higher_orders}
{\cal V}_{
eff,N}(B,E)&=&-\frac{E^{2N} B e^{2N+1}}{8\pi^2}
\nonumber \\
&\times&
\int_0^\infty ds \exp(-m^2s)\lambda_N s^{2N-2} 
\coth(e s B)\nonumber\\
&=&-\frac{E^{2N} e^{2}}{B^{2N-2}8\pi^2} \lambda_N 
\nonumber \\
&\times&
\int_0^\infty dx 
\exp\left(-\frac{m^2}{eB}x\right)x^{2N-2} \coth(x)\nonumber\\
&=&\frac{E^{2N} e^{2}}{B^{2N-2}8\pi^2} \lambda_N \left(\frac{1}{2^{2N-2}} 
\Psi^{(2N-2)}\left(\frac{m^2}{2eB}\right) \right.
\nonumber \\
&+&\left.
\frac{(2N-2)!}{\left(\frac{m^2}{eB}\right)^{2N-1}}\right)
\end{eqnarray}
with a numerical factor $\lambda_N$ which is given in 
Table \ref{tab:Coefficients} for $N$ up to 8. 
It decreases approximately exponentially with the order of the 
expansion N. To get the corresponding terms in the energy density one has to multiply the ${\cal V}_{eff,N}$ by an additional factor of $2N-1$, of course, due to Eq.(\ref{eq:energy_density}). The result can be tested again using the Euler-Heisenberg Lagrangian. For $N=2$ and small magnetic fields it reproduces exactly the missing term of order $E^4$ in Eq.(\ref{eq:euler_heisenberg})
\begin{eqnarray}
{\cal V}_{eff,2}(B,E)&=&\frac{E^{4} e^{2}}{B^{2}8\pi^2} \left(-\frac{1}{45}\right) \left(\frac{1}{4} \Psi^{(2)}\left(\frac{m^2}{2eB}\right)+\frac{2}{\left(\frac{m^2}{eB}\right)^{3}}\right)\nonumber\\
&=&\frac{E^4 e^4}{360 \pi^2 m^4}+E^4 O(B^2).
\end{eqnarray}

\begin{table*}
\caption{Coefficients of the expansion to higher orders in the electric field}
\begin{center}
\begin{tabular}{c|c|c|c|c|c|c|c}
		 $N$	& 2&3&4&5&6&7&8\\
		 \hline
			$\lambda_N$& 	$-\frac{1}{45}$&$-\frac{2}{945}$&$-\frac{1}{4725}$&$-\frac{2}{93555}$&$-\frac{1382}{638512875}$&$-\frac{4}{18243225}$&$-\frac{3617}{162820783125}$\\
		\end{tabular}
\end{center}
\label{tab:Coefficients}
\end{table*}

\begin{figure}
	\centering
		\includegraphics[width=0.45\textwidth]{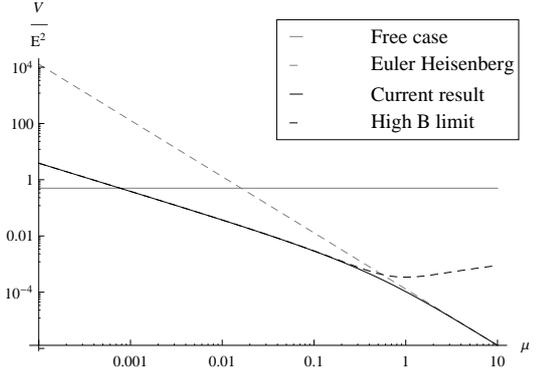}
	\caption{Relative magnitudes of relevant terms in the lagrangian which are quadratic in the electric field as a function of $\mu=m^2/(eB)$; $\mu$ is in heavy ion collisions of the order of $10^{-4}$ to $10^{-5}$.}
	\label{fig:loglogplotpotentials}
\end{figure}

In heavy ion collisions $B$ is much larger than $m_e$ 
such that one can use the asymptotic expansion of the $\psi$ function
\begin{equation}
{\cal V}_{eff}(B,E)~=~\frac{\alpha E^2}{6\pi} \left[ 
\frac{e B}{m^2} +\gamma+\ln\frac{m^2}{eB} \right].
\end{equation}
For $B=(100 {\rm MeV})^2$ one finds ${\cal V}_{eff}(B,E)\sim 4.5 E^2$,
nine times the free electric field energy. Plots of the relevant quantities are shown in Fig. \ref{fig:loglogplotpotentials}.
As argued in \cite{ba} this
increase of the total electric field energy by a factor 10 will in turn 
reduce the induced electric field strength substantially. 
Even more important is the fact that  ${\cal V}_{eff}$ grows linearly in 
$B$, just as the driving topological term. If it had grown 
faster (e.g. like $B^2$ as for the Euler-Heisenberg Lagrangian), the
CME would have died out with increasing $B$. As it is the induced 
electric field is only little $B$ dependent for strong $B$ fields.    
However, one
should keep in mind that our result applies only to the situation 
of constant fields, which is not a good approximation for a 
heavy ion collision.

\section{An alternative calculation using the explicit form of the fermion propagator}

Schwinger's approach is the most efficient one, when the aim is  
to derive the effective higher-order photon action. However, one often wants 
to study dynamical quantum processes in a constant background field.
To do so, one needs the modified Feynman rules in such a background field. 
In this section we will try to re-derive the result just obtained for
$\Theta=0$ with 
the Schwinger approach, namely the effective action in second order in 
the electric field and all orders in the magnetic one, using the explicit 
form of the fermion Feynman propagator given in \cite{ruder}. This exercise
is meant to demonstrate that one can actually do so, but also that special 
care is needed with respect to finite renormalisation terms. The propagator 
reads
\begin{eqnarray}
\label{S}
i S(\chi',\chi)&=&e^{-i \frac{e B}{2}(x'+x)(y'-y)}\frac{1}{8 \pi^2 i}\int_0^\infty d\mu
\nonumber \\
&\times&
\frac{e B}{2 (\mu+{\rm i} \epsilon) \sin\left(\frac{e B}{2(\mu+{\rm i} \epsilon)}\right)} 
e^{-\frac{i m^2}{2(\mu+{\rm i} \epsilon)}}
\nonumber \\ 
&\times&
e^{\frac{i e B}{4} \cot\left(\frac{e B}{2(\mu+{\rm i} \epsilon)}\right)\left((x'-x)^2+(y'-y)^2\right)}\nonumber\\
&\times&
e^{\frac{i (\mu+{\rm i}\epsilon)}{2}\left((z'-z)^2-(t'-t)^2\right)}M 
e^{-i\sigma^{12} \frac{e B}{2 (\mu+{\rm i}\epsilon)}}\\
M&=&\gamma^0(\mu+{\rm i}\epsilon)(t'-t)-\gamma^3(\mu+{\rm i}\epsilon)(z'-z)\nonumber\\
&+&\frac{eB}{2}\left(\gamma^1(y'-y)-\gamma^2(x'-x)\right)\nonumber\\
&-&\frac{eB}{2}\cot\left(\frac{eB}{2(\mu+{\rm i}\epsilon)}\right)\left(\gamma^1(x'-x)+\gamma^2(y'-y)\right),\nonumber \\
\end{eqnarray}
where an $\epsilon$ prescription was introduced such as to make a Wick rotation 
in $\mu$ possible, which lies at the heart of this approach. 
It will always make the results finite 
and thus includes implicitly a regularisation. As mentioned in the 
Introduction different 
renormalisation schemes differ by finite renormalisation terms. 
In the Gepr\"ags et al. approach 
this ambiguity might show up as an ambiguity in the choice of the $\epsilon$ 
prescription, but this was not explored so far. To simplify the notation the $\epsilon$ will be suppressed from now on.\\
The situation is rather complicated.  The $e^{i\cot{\frac{1}{z}}}$ function has a countably infinite number of essential singularities on the positive real axis at
\begin{eqnarray}
	z_N=\frac{1}{N \pi}, \forall N\in I\!\! N.
\end{eqnarray}
Thus the analytic continuation is highly non-trivial and 
we will show that for our case this prescription leads to a different 
finite renormalisation than the Schwinger formula. (The leading 
logarithmic divergence is naturally renormalised in the same way.)
Physics-wise the problem can be linked to the existence of Landau orbitals in a 
constant magnetic field. The Fourier transform of Eq.(\ref{S}) gives
typically Gaussians of the form:
\begin{equation}
\exp\{ {\rm i} (p_x^2+p_y^2)\tan(eB/2\mu)/(eB)\}
\end{equation}
and for e.g. $\tan(eB/2\mu)=\pi$ one gets identical weight factors
for all Landau orbits with 
\begin{eqnarray}
E~=~ s_z\frac{a_feB}{m} &+& \sqrt{m^2+p_z^2+(2n+2s_z+1)m\omega_c} 
\nonumber \\
\omega_c &=& \frac{eB}{m}\\
p_x^2+p_y^2 &=& (2n+2s_z+1)eB
\end{eqnarray}
with the anomalous magnetic moment $a_f$. This can also be interpreted 
more intuitively as follows: UV regularisation is concerned with the 
short distance behavior. For arbitrarily small $eB>0$ all classical Landau 
orbits return to the starting point and thus give an unsuppressed contribution. 
In contrast Schwinger subtracts only the $B=0$ contribution.
One cannot expect that the $\mu$ integrals from Eq.(\ref{S}), the sum over 
Landau orbits and the limit $B\rightarrow 0$ commute. 
The Gepr\"ags et al. result should still give the correct answer,
but only up to finite renormalization terms.
Therefore, we will 
proceed as follows: we will apply this approach disregarding the
problems just discussed 
and will adjust the finite regularisation terms in the final 
expression to the Schwinger result. (We were not able to find an $\epsilon$ 
prescription which would automatically result in an expression in the 
same renormalisation scheme as Schwinger's approach.)
Luckily, the changes to be made are rather obvious.  
In any case, we find it very reassuring that 
up to this finite regularisation term both approaches lead to 
the same result.\\

\begin{figure}
	\centering
		\includegraphics[width=0.30\textwidth]{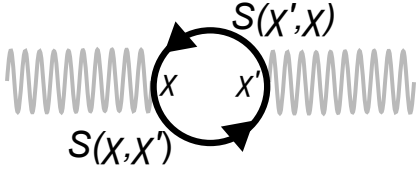}
	\caption{Sketch of the vacuum polarization Feynman diagram.}
	\label{fig:graph}
\end{figure}

We treat the magnetic field $\vec B$ exactly, i.e. as part of ${\cal H}_0$ 
and the electric field as perturbation, i.e. as part of 
${\cal H}_{int}$ and equate the second order of the expectation value 
\begin{equation}
\label{vac}
\langle\vec B|T\{e^{i\int d^4\chi {\cal L}_{int}(\chi)}\}|\vec B\rangle
\end{equation}
with the first order of the effective electromagnetic Lagrangian 
we are seeking
(which is of second order in the electric field $\vec E$), i.e.
\begin{equation}
{\cal L}_{eff}(\chi_1) =\langle\vec B|T\left\{\frac{i}{2}\int d^4\chi_2 {\cal L}_{int}(\chi_1){\cal L}_{int}(\chi_2)\right\}|\vec B\rangle.
\end{equation}
We basically calculate the vacuum polarization graph depicted in Fig. \ref{fig:graph}
for a constant $\vec E$ field in the background of an 
external constant $\vec B$ field. 
For simplicity we only discuss the case $\vec E=E\vec e_z$, 
$\vec B=B\vec e_z$ most relevant for the CME. \\
For constant fields, rather than 
point photons, the propagator from $\chi_1$ to $\chi_2$ fulfills 
\begin{equation}  
\Bigl[ {\rm i} \gamma^{\mu} {\cal D}_{\mu} -m \Bigr] S(\chi_1,\chi_2) ~=~
\delta^4(\chi_1-\chi_2) 
\end{equation}
with the gauge invariant derivative rather than the usual one. This
introduces an overall gauge factor, which in turn leads to 
covariant derivatives of the form 
\begin{equation}  
{\cal D}_{\mu} ... ~=~ 
\Bigl(\partial_{\mu} -\frac{eQ}{2}F_{\mu\nu}(x-x')^{\nu}\Bigr) ...
\end{equation}
see \cite{gauge_inv}. In \cite{Nam:2009jb} this was actually already 
discussed in connection with the CME.\\
We chose the 4-coordinates 
$(\chi^\mu_1):=\left(t,x,y,z\right)$,
$(\chi^\mu_2):=\left(t',x',y',z'\right)$.
The expression to be calculated is
\begin{eqnarray}
\label{Seff}
{\cal L}_{eff} &=& \frac{i}{2}\int d^4\chi_2~E^2(z-z')^2
\Pi^{00}(\chi_1,\chi_2)
\\
\Pi^{\mu\nu}&=&-e^2Tr(\gamma^\nu S(\chi_1,\chi_2)\gamma^\mu S(\chi_2,\chi_1)),
\end{eqnarray}
where we adopted the gauge $(A^\mu)(\chi):=\left(zE,0,0,0\right)$.

With the abbreviations
\begin{eqnarray}
\label{mmu}
m_0&:=&\mu(t'-t)\\
m_1&:=&\frac{eB}{2}\left((y'-y)-\cot\nu(x'-x)\right)\nonumber\\
m_2&:=&\frac{eB}{2}\left(-(x'-x)-\cot\nu(y'-y)\right)\nonumber\\
m_3&:=&-\mu(z'-z)\nonumber\\
\label{nu}
\nu&:=&\frac{eB}{2\mu}
\end{eqnarray}
one gets 
\begin{eqnarray}
M &=& m_\mu\gamma^\mu\\
\label{pi}
\Pi^{00}&=&e^2\left(\frac{1}{8 \pi^2 i}\right)^2\int_0^\infty d\mu\frac{\nu}{\sin\nu}
	\int_0^\infty d\tilde{\mu}\frac{\tilde\nu}{\sin\tilde\nu} e^{ -\frac{i m^2}{2}\left(\frac{1}{\mu}+\frac{1}{\tilde\mu}\right) }
\nonumber \\ 
&\times&e^{\frac{i e B}{4} (\cot\nu + \cot\tilde\nu)\left((x'-x)^2+(y'-y)^2\right)}
\nonumber \\
&\times&e^{\frac{i (\mu+\tilde\mu)}{2}\left((z'-z)^2-(t'-t)^2\right)}
\nonumber\\
&\times&
Tr(\gamma^0 M e^{-i\sigma^{12} \nu}\gamma^0 \tilde M e^{-i\sigma^{12} \tilde\nu}).
\end{eqnarray}
Note that the first exponentials of the propagators cancelled in Eq. (\ref{pi}) so that 
$\Pi^{00}(\chi',\chi)\equiv\Pi^{00}(\chi'-\chi)$. \\
Evaluating the exponentials in the trace yields
\begin{equation}
e^{-i\sigma^{12} \nu}~ =~ \cos\nu-i\sigma^{12}\sin\nu.
\end{equation}
The  evaluation of the trace is straight forward and gives
\begin{eqnarray}
\Pi^{00}(\Delta)&=&\left(\frac{e}{4 \pi^2}\right)^2\int_0^\infty 
d\mu \int_0^\infty d\tilde{\mu}\frac{\nu}{\sin\nu}\frac{\tilde\nu}
{\sin\tilde\nu} 
\nonumber \\
&\times& e^{\frac{i e B}{4} (\cot\nu + \cot\tilde\nu)\left(\Delta x^2+\Delta y^2\right)}
e^{\frac{i (\mu+\tilde\mu)}{2}\left(\Delta z^2-\Delta t^2\right)}\nonumber\\
&\times&e^{ -\frac{i m^2}{2}\left(\frac{1}{\mu}+\frac{1}{\tilde\mu}\right) } \left(\cos(\tilde\nu+\nu)\mu\tilde\mu(\Delta t^2+\Delta z^2)
\right.
\nonumber \\
&-&\left.\left(\frac{eB}{2}\right)^2(\Delta x^2+\Delta y^2)\frac{1}
{\sin\nu\sin\tilde\nu}\right)
\end{eqnarray}
with the notation $\chi_1-\chi_2=(\Delta t,\Delta x,\Delta y,\Delta z)=\Delta$
and $\chi= (\chi_1+\chi_2)/2$. Now we rotate the $\mu$ and $\tilde\mu$ integrals 
to the positive imaginary axis, and perform a Wick-rotation for $\Delta t$,
yielding
\begin{eqnarray}
{\cal L}_{eff} &=&
\left(\frac{eE}{4 \pi^2}\right)^2\int_0^\infty d\mu \int_0^\infty d\tilde{\mu}
\frac{\nu}{\sinh\nu}\frac{\tilde\nu}{\sinh\tilde\nu} 
\nonumber \\
&\times&
e^{ -\frac{m^2}{2}\left(\frac{1}{\mu}+\frac{1}{\tilde\mu}\right) }
\nonumber\\
&\times&\int d^4\Delta \Delta z^2e^{-\frac{e B}{4} 
(\coth\nu + \coth\tilde\nu)\left(\Delta x^2+\Delta y^2\right)}
\nonumber \\
&\times&
e^{-\frac{(\mu+\tilde\mu)}{2}\left(\Delta z^2+\Delta t^2\right)} \nonumber
\\
&\times&\left(\cosh(\tilde\nu+\nu)\mu\tilde\mu(\Delta z^2-\Delta t^2)\right.
\nonumber \\
&-&\left.\left(\frac{eB}{2}\right)^2(\Delta x^2+\Delta y^2)\frac{1}{\sinh\nu\sinh\tilde\nu}\right).
\end{eqnarray}
After performing all Gaussian integrals 
this simplifies to 
\begin{eqnarray}
{\cal L}_{eff}&=&\frac{e^2E^2}{4 \pi^2}\int_0^\infty d\mu \int_0^\infty 
d\tilde{\mu}\left(\frac{2}{eB}\right)^2
e^{ -\frac{m^2}{2}\left(\frac{1}{\mu}+\frac{1}{\tilde\mu}\right) }
\\
&\times&\frac{\nu^3\tilde\nu^3}{(\nu+\tilde\nu)^2\sinh(\nu+\tilde\nu)} 
\left( \frac{\cosh(\nu+\tilde\nu)}{\nu+\tilde\nu} -\frac{1}{\sinh(\nu+\tilde\nu)} \right),\nonumber
\end{eqnarray}
which is an even function in $B$, allowing us to substitute $\mu$, 
$\tilde \mu$ by $\rho=|\nu|$, $\tilde\rho=|\tilde\nu|$. In terms of 
these variables we introduce $\sigma=\rho+\tilde \rho$ and 
$\delta=\rho-\tilde \rho$ and perform the $\delta$ integration to obtain 
\begin{eqnarray}
{\cal L}_{eff}&=&\frac{e^2E^2}{8 \pi^2}
\int_0^\infty d\sigma\int_{-\sigma}^\sigma d\delta
e^{-\frac{m^2}{e\left|B\right|}\sigma} \frac{\sigma^2-\delta^2}{4\sigma^2} 
\nonumber \\
&\times&
\left( \frac{\coth\sigma}{\sigma} -\frac{1}{\sinh^2\sigma} \right)\\
&=&\frac{e^2E^2}{24 \pi^2}\int_0^\infty d\sigma
e^{-\frac{m^2}{e\left|B\right|}\sigma} \left( \coth\sigma 
-\frac{\sigma}{\sinh^2\sigma} \right),\nonumber
\end{eqnarray}
which differs from the result obtained in the last section  
\begin{eqnarray}
{\cal L}_{eff}^{Schwinger}&=&\frac{e^2E^2}{24 \pi^2}\int_0^\infty d\sigma
e^{-\frac{m^2}{e\left|B\right|}\sigma} \left( \coth\sigma -\frac{1}{\sigma} 
\right)
\end{eqnarray}
only in the subtracted renormalisation term. While in the Schwinger treatment 
the latter is $B$ independent
\begin{equation} 
\int_0^\infty d\sigma
e^{-\frac{m^2}{e\left|B\right|}\sigma} \frac{1}{\sigma}~=~ 
\int_0^\infty d\sigma
e^{-{\sigma}}\frac{1}{\sigma} 
\end{equation}
the former is not.
We correct this by simply adding the relative finite renormalisation 
term
\begin{eqnarray}
\Delta{\cal L}_{eff}^{Schwinger}&=&\frac{e^2E^2}{24 \pi^2}\int_0^\infty d\sigma
e^{-\frac{m^2}{e\left|B\right|}\sigma} \left(\frac{1}{\sigma}-
\frac{\sigma}{\sinh^2\sigma}\right).\nonumber \\
\end{eqnarray}
We conclude that the approach from Gepr\"ags et al. can be used equally well, 
if the renormalisation for all loop graphs is adapted to the usual conventions.

\section{Conclusions}

In this publication we have calculated higher-order terms of the effective 
electromagnetic Lagrangian. While strong QED is an 
interesting and active field in its own right, with applications in 
e.g. the astrophysics of magnetars, these studies were specifically 
motivated by the Chiral Magnetic Effects (CME), possibly observed in 
high-energy heavy ion collisions. The postulated mechanism is that the extremely 
strong magnetic fields present in the early phase of such a collision in 
combination with topological tunneling in QCD could induce an electric field,
subsequently generating specific charge correlations. For any such mechanism
strong QED effects are so large that they have to be taken into account
for any quantitative description. This was demonstrated in our paper for
the case of constant electric and magnetic fields. The observed charge 
correlations are small, hinting to an electric field strength of the order
100 MeV$^2$, while the magnetic field strength is of the order $10^4 \mbox{ MeV}^2$,
such that the most relevant term is of second order in $E$ and all orders in 
$B$. We calculated this contribution.\\
The result we got implies that strong QED effects increase the energy density 
associated with an electric field by an order of magnitude for 
$B=(100 \mbox{ MeV})^2$, thus strongly suppressing the size of such fields as 
compared to naive expectations. This result is obviously most relevant
for the CME, although the approximation of constant fields is probably a
bad one for the heavy-ion setting where the fields change on time and 
distance scales of several fm, which is also the radius of Landau orbits 
for the $B$ fields considered. If the CME is confirmed by future experiments 
at RHIC and especially LHC, where the fields will still be stronger and 
even more Lorentz contracted, one will have to develop techniques to
treat also the dynamics of the strong  QED effects reliably.\\
For calculations  of e.g. Compton scattering in the accretion column of a 
magnetar one needs the Feynman rules for a magnetic background field, as 
given e.g. by Gepr\"ags et al. \cite{ruder}. Therefore, we performed 
our calculation also with these explicit Feynman rules and showed that  
the result agrees with that calculated in the Schwinger approach up
to a finite renormalisation term. We see this as an important check 
for the Gepr\"ags et al. approach, which also illustrates nicely some of 
the technical problems encountered by any such calculation.\\ 
Finally, in the Schwinger approach we generalized the result to
higher orders in the $E$ field strength.\\

\section{Acknowledgement}

This work was supported by BMBF. A.S. also acknowledges 
the hospitality of the Yukawa Institute, Kyoto, Japan, 
where part of this work was done. We thank Berndt M\"uller and 
Kenji Fukushima for fruitful discussions.


\end{document}